# Dynamic Difficulty Adjustment With Brain Waves as a Tool for Optimizing Engagement

Nir Cafri[a]

The Departments of Physiology & Cell Biology and Cognitive & Brain Sciences, Zlotowski Center for Neuroscience, Ben-Gurion University of the Negev, Beer-Sheva, Israel[a]

[a]nircaf@post.bgu.ac.il

*Abstract*— **Dynamic difficulty adjustment (DDA) is a mechanism that allows tailored gaming experience according to individual player's characteristics. This study focuses on the engagement effect due to DDA in Virtual Reality (VR). DDA with input from Electroencephalography (EEG) is used to measure engagement in a VR game. An intra-subject study of DDA vs non-DDA engagement periods showed a 19.79% increase in engagement period in the DDA session.**

## I. Introduction

Electroencephalography (EEG) is a noninvasive method for investigating brain activity. It measures the change in brain waves activity in real time and for personal use[1]. The frequency bands are; $\delta$ (1–4 Hz), $\theta$ (5–8 Hz), $\alpha$ (9–13 Hz), $\beta$ (12–30 Hz) and $\gamma$ (30–50 Hz).

Engagement is a state of optimal experience that is characterized by focused attention, clear mind, mind and body unison, effortless concentration, loss of self-consciousness, distortion of time, and intrinsic enjoyment [2].

Task Engagement Index (TEI) is defined as the level of effortful concentration and striving towards task goals where task demands and personal characteristics may influence the pattern of processing[3]. TEI is measured from EEG input as $\beta/(\alpha + \theta)$[4-5].

Dynamic Difficulty Adjustment (DDA) is a method of automatically modifying a game's features, behaviors, and scenarios in real-time, depending on the player's skill, so that the player does not feel bored when the game is very simple or frustrated when the game is difficult[6-9].

Head-Mounted Display (HMD) is a form of display generated by a computer software which allows virtual reality (VR) to simulate a real environment that is then experienced by a user through a human–machine interface[10-11].

Previous studies showed that DDA and Functional near-infrared spectroscopy can optimize workload[12], DDA and EEG in personal computer video games can increase engagement[13], and EEG can estimate feelings in VR [14]. This study shows a combination of EEG, DDA and VR, hypothesizing that DDA will increase engagement period.

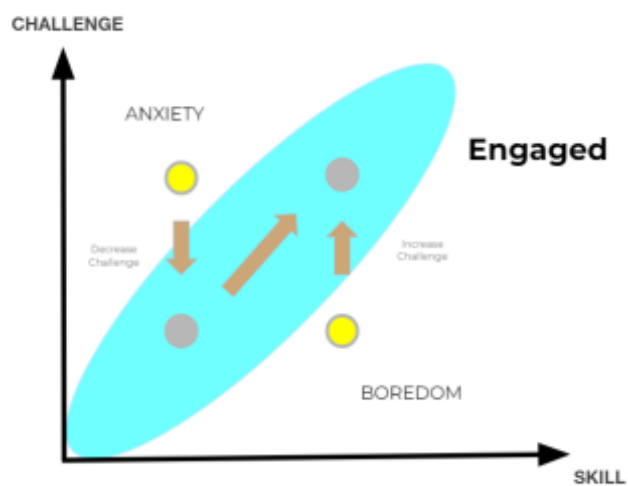

Fig. 1. Engagement theory

## II. Methods

TEI was calculated from EEG input that was measured with MUSE S EEG system (InterAxon Inc.)[15] from the average of frontal lobe electrodes: Fp1 and Fp2[16]. The HMD platform used is Oculus Quest 2 (Meta Inc.)[17].

Score, health and death counts were displayed as a motivation for the participants to optimize interest, immersiveness, and to keep track of game progress[SV1].

The participants went through two sessions that were randomly ordered: A (Control; Non- DDA)

and B (DDA). Session A involved 6 minutes playing while enemies respawned every 15 seconds. B was broken into three sub-sessions: B.1 involved 3 minutes setting a low threshold and no enemies spawned. B.2 involved 3 minutes setting a high threshold and enemies spawned every 5 seconds. B.3 involved 6 minutes playing with enemies spawning when TEI was below low threshold -boredom (fig.1) and enemies disappearing when TEI was above high threshold-anxiety (fig.1).

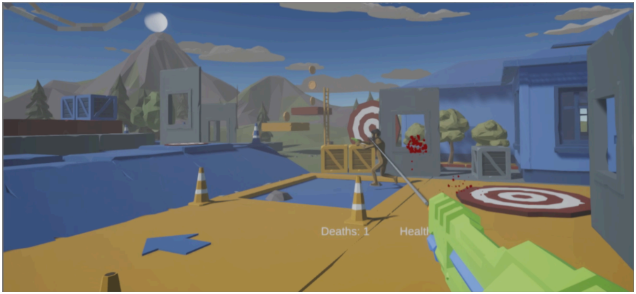

Fig. 2. Gameplay example - ball of light up the mountain is set as an in-game DDA adjustment indicator.

Participants started with 3 minutes getting to know the environment, followed by the two study sections (A and B) in a random order.

Percent of time engaged was calculated by the time TEI values were between low and high thresholds, divided by total time[18]. Comparison was made between the 6 minutes of Non-DDA and the 6 minutes of DDA sessions to determine engagement success. N= 6 participants in the study: age= 31.83(±2.54), gender= 50%.

### III. RESULTS

Percent of time engaged in the DDA session(fig 3) was 71.0% (±8.07%). Percent of time engaged in the no-DDA session was 51.2% (±5.84%). Mann Whitney U test P value =0.008. Effect size Cohen's d = 2.513. N=6.

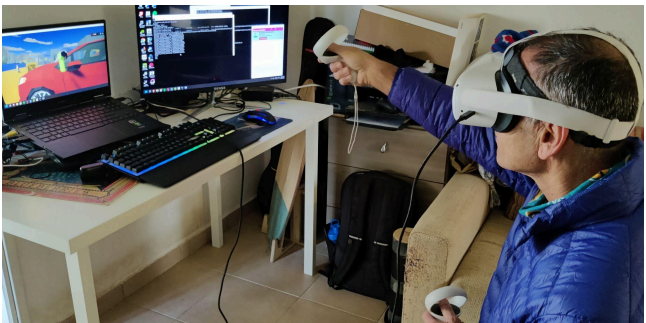

Fig. 3. Participant's during the experiment wearing the Quest 2 and Muse S.

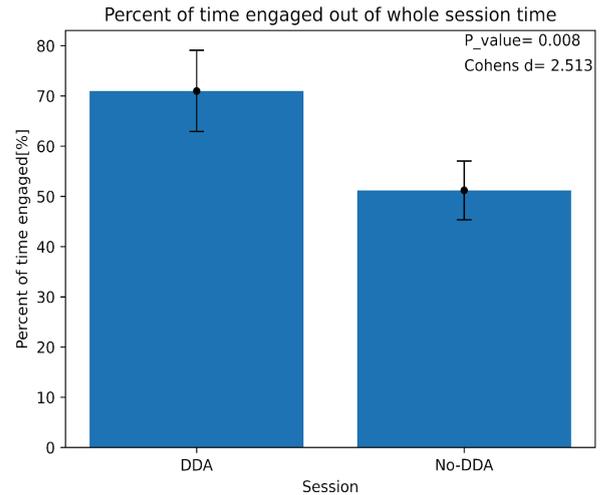

Fig. 4. Percent of time engaged in DDA and no-DDA sessions.

### IV. DISCUSSION

The participants showed an increase in percent of time engaged while DDA was active. This increase of 19.79%(±2.22%) on average confirms the hypothesis that DDA increases engagement period.

However, the small sample size and the short trial weakens the conclusions that can be drawn. There could be a number of factors that may have affected the results such as apathy to a familiar platform or game, or insufficient challenge in the Non-DDA session.

The usage of consumer grade devices in this study (both the EEG and VR were below $300) allows various applications of these findings to fields such as physical therapy, adaptive learning, entertainment, and workflow optimization.

Further studies could apply these findings to: other game genres (e.g. RPG, RTS, MOBA), bigger sample size, verification of EEG signal with other HMDs, use of machine learning to make more accurate TE, and Incorporation of additional biomarkers (e.g. GSR, HR) can be implemented.

ACKNOWLEDGMENT

Assets: Oculus Integration, POLYGON Starter Pack, Simple FX - Cartoon Particles, GameDevChef BlueMuse, MuseLSL. Supervisor: Shahar Maidenbaum. Subjects: Alon, Gidon, Maayan, Dina, Ellie and Nir. Unity consulting: Yaniv. Equipment: Oleg.